\documentclass[12pt,aps,floats,prci,tightenlines,showpacs]{revtex4}
\usepackage{amsmath}  
\usepackage{amsthm}  
\usepackage{amsfonts}  
\usepackage{amssymb} 
\usepackage{delarray} 
\usepackage{graphicx}
\usepackage{dcolumn}  
\usepackage{epsfig}  
\usepackage{float}
\restylefloat{figure}
\restylefloat{table}

\begin{document}

\title{Nuclear matter in the crust of neutron stars}
\author{P. G\"ogelein and H. M\"uther}
\affiliation{Institut f\"ur Theoretische Physik, \\
Universit\"at T\"ubingen, D-72076 T\"ubingen, Germany}

\vskip .3cm


\vskip .3cm

\begin{abstract}
The properties of inhomogeneous nuclear matter are investigated considering the
self-consistent Skyrme Hartree-Fock approach with inclusion of pairing
correlations. For a comparison we also consider a relativistic mean field
approach. The inhomogeneous infinite matter is described in terms of cubic
Wigner-Seitz cells, which leads to a smooth transition to the limit of
homogeneous nuclear matter. The possible existence of various structures in
the so-called pasta phase is investigated within this self-consistent approach and
a comparison is made to results obtained within the Thomas-Fermi approximation.
Results for the proton abundances and the pairing properties are discussed for
densities for which clustering phenomena are obtained.
\end{abstract}

\maketitle

\section{Introduction}
The crust of neutron stars is a very intriguing object for theoretical nuclear
structure physics, as it contains the transition from stable nuclei in the outer
crust to a system of homogeneous nuclear matter, consisting of protons, neutrons
and leptons in $\beta$-equilibrium, in the inner part of this crust. The
question of how matter consisting of isolated nuclei melts into uniform matter
with increasing density has evoked a large number of
studies\cite{raven1, hashimo, oyama1, oyama2}. Already at moderate densities the
Fermi energy of the electron is so high that the $\beta$-stability enhances the
neutron fraction of the baryons so much that a part of these neutrons drip out
of the nuclei. This leads to a structure, in which quasi-nuclei, clusters of
protons and neutrons, are embedded in a sea of neutrons. In order to minimise
the Coulomb repulsion between the protons, the quasi-nuclei form a lattice.

Therefore one typically describes these structures in form of the Wigner-Seitz
(WS) cell approximation. One assumes a geometrical shape for the quasi-nuclei and
determines the nuclear contribution to the energy of such a WS cell from a
phenomenological energy-density functional. Such Thomas-Fermi calculations yield
a variety of structures: Spherical quasi-nuclei, which are favoured at small
densities, merge with increasing density to strings, which then may cluster to
parallel plates and so on. These geometrical structures have been the origin for
the popular name of this phase: Pasta phase.         

Such Thomas-Fermi calculations, however, are very sensitive to the surface tension
under consideration. Furthermore they do not account for characteristic
features of the structure of finite nuclei, like the shell-effects. Shell
effects favour the formation of closed shell systems and may have a significant
effect on the formation of inhomogeneous nuclear structures in the crust of neutron
stars. These shell effects are incorporated in self-consistent Hartree-Fock or mean
field calculations, which can treat finite nuclei, infinite matter and
inhomogeneous structures in between within a consistent frame based on
an effective nucleon-nucleon interaction. Such calculations employing the
density-dependent Skyrme forces\cite{sk1,sk2} have been done more than 25 years 
ago by Bonche and Vautherin\cite{bv81} and by a few other groups. 

These studies show indeed that shell effects have a significant influence on
details like the proton fraction of the baryonic matter in the inhomogeneous 
phase\cite{PC:Montani04}. They also provide the basis for a microscopic
investigation of properties beyond the equation of state. This includes the study
of pairing phenomena, excitation modes and response functions as well as the
effects of finite temperature.

Self-consistent Hartree-Fock calculations for such inhomogeneous nuclear
structures have typically been performed assuming a WS cell of spherical shape.
This assumption of quasi-nuclei with spherical symmetry reduces the numerical
work-load considerably. However, it does not allow the exploration of
quasi-nuclear clusters in form of strings or plates as predicted from Thomas-Fermi
calculations. Furthermore the limit of homogeneous matter can not be described in a
satisfactory manner in such a spherical WS cell. Employing the representation of 
plane wave single-particle states in terms of spherical Bessel functions leads to
a density profile, which, depending on the boundary condition chosen, exhibits
either a minimum or a maximum at the boundary of the cell. Bounche and
Vautherin\cite{bv81} therefore suggested to use a mixed basis, for which, depending
on the angular momentum, different boundary conditions were considered. However,
even this optimised choice leads to density profiles with
fluctuations\cite{PC:Montani04}. 

Therefore the investigations presented here consider cubic WS cells, which allows
for the description of non-spherical quasi-nuclear structures and contains the
limit of homogeneous matter in a natural way. Self-consistent Hartree-Fock
calculations are performed for $\beta$-stable matter at densities for which the
quasi-nuclear structures discussed above are expected. For the nuclear Hamiltonian
we consider various Skyrme forces but also perform calculations within the
effective relativistic mean-field approximation. Special attention will be paid to
the comparison between results obtained in the Hartree-Fock approach and 
corresponding Thomas-Fermi calculations.

After this introduction we will briefly review the Hartree-Fock approximation
using Skyrme interactions and the technique used to solve the equations
resulting from this approach employing the imaginary time step method in section
2. We then turn to the relativistic mean field approach and the adaption of the
imaginary time step method to be used within this relativistic framework. After
a short description on the inclusion of pairing correlations in section 4, we
present results in section 5. The main conclusions are summarised in the final
section 6.

\section{Skyrme--Hartree--Fock Calculations}

\subsection{Energy Functional}
The Skyrme--Hartree--Fock approach has frequently been described in the
literature\cite{sk1,sk2,bv81,NMB:Ring80}. Therefore we will restrict the
presentation here to a few basic equations, which will define the nomenclature.
The Skyrme model is defined in terms of an energy density $\mathcal{H} 
(\boldsymbol{r})$, which can be split into various 
contributions\cite{sk2,SP:Chabanat98}
\begin{equation}
 \mathcal{H} = 	\mathcal{H}_K + \mathcal{H}_0 + \mathcal{H}_3 
 		+ \mathcal{H}_{\text{eff}} + \mathcal{H}_{\text{fin}} 
		+ \mathcal{H}_{\text{so}} 
		+ \mathcal{H}_{\text{Coul}}, \label{eq:sk1}
\end{equation}
where $ \mathcal{H}_K $ is the kinetic energy term,  
$ \mathcal{H}_0 $ a zero range term, $ \mathcal{H}_3 $ a density dependent term, 
$\mathcal{H}_{\text{eff}} $ an effective mass term, $ \mathcal{H}_{\text{fin}} $
a finite range term and $\mathcal{H}_{\text{so}} $ a spin-orbit term. These
terms are given by
\begin{eqnarray}
\mathcal{H}_K 	
	& = & \frac{ \hbar^2}{2m} \tau,  \notag \\
\mathcal{H}_0 	
	& = & \textstyle{\frac{1}{4}} t_0 
	      \big[ (2+x_0) \rho^2 - ( 2x_0 + 1 ) (\rho_p^2 + \rho_n^2 ) \big],  \notag \\
\mathcal{H}_3 	
	& = & \textstyle{\frac{1}{24}} t_3 \rho^\alpha 
	      \big[ (2 + x_3 ) \rho^2 - ( 2x_3 + 1 ) ( \rho_p^2 + \rho_n^2 ) \big], \notag \\
\mathcal{H}_{\text{eff}}	
	& = & \textstyle{\frac{1}{8}} 
		\big[ t_1 (2 + x_1 ) + t_2 (2 + x_2 ) \big] \tau \rho \notag\\
	&   & + \textstyle{\frac{1}{8}} \big[ t_2 ( 2x_2 +1 ) - t_1 ( 2 x_1 + 1 ) \big] 
		\big[ \tau_p \rho_p + \tau_n \rho_n \big], \notag \\
\mathcal{H}_{\text{fin}} 	
	& = & -\textstyle{\frac{1}{32}} \big[3 t_1 ( 2 + x_1 ) - t_2 (2 + x_2 ) \big] 
		\rho \Delta \rho \notag \\
	& & + \textstyle{\frac{1}{32}} \big[ 3t_1(2x_1 + 1) + t_2(2x_2 + 1 ) \big]
		\big[\rho_p \Delta \rho_p  +\rho_n \Delta \rho_n  \big], \notag \\
\mathcal{H}_{\text{so}} 	
	& = & - \textstyle{\frac{1}{2}} W_0 
	\big[ \rho \, \boldsymbol{\nabla}\boldsymbol{J} 
	      +  \rho_p \, \boldsymbol{\nabla}\boldsymbol{J}_p 
	      +  \rho_n \, \boldsymbol{\nabla}\boldsymbol{J}_n  \big].
	      \label{eq:sk2}
\end{eqnarray}
The coefficients $t_i$, $x_i$, $W_0$, and $\alpha$ are the parameters of a 
generalised Skyrme force\cite{TD:Bonche85}. The energy density of
eq.(\ref{eq:sk1}) contains furthermore the contribution of the Coulomb force, 
$\mathcal{H}_{\text{Coul}}$, which 
is calculated from the charge density $\rho_C$ as
\begin{equation}
\mathcal{H}_{\text{Coul}}
	= \frac{e^2}{2}\rho_C (\boldsymbol{r}) \int d^3r' \,
		\frac{\rho_C (\boldsymbol{r'}) }{| \boldsymbol{r} \,-\, 
		\boldsymbol{r'} |} \
	  - \frac{3 e^2}{4}  \left( \frac{3}{\pi} \right) ^{1/3} \rho_C^{4/3}. 
\end{equation}
Here the exchange part of the Coulomb term is calculated within the Slater 
approximation. Following \cite{TD:Bonche85}
the center-of-mass recoil energy has been approximated as 
$-\sum \boldsymbol{p}_i^2/2Am$.

The densities $\rho$, $\tau$, and $\boldsymbol{J}$ are defined in terms of the
corresponding densities for protons and neutrons 
$\rho = \rho_p + \rho_n$, $\tau_p + \tau_n$ 
and \mbox{$\boldsymbol{J} = \boldsymbol{J}_p + \boldsymbol{J}_n$}. If we
identify the isospin label $(q = n, p)$, the corresponding matter densities are
given by
\begin{equation}
\rho_q ( \boldsymbol{r} ) 
  = \sum_{k,s}\ \eta_k^q \, |\varphi_k^q (\boldsymbol{r}, s ) |^2,
\end{equation} 
where $\varphi_k^{q} (\boldsymbol{r},s)$ is the single-particle wave function 
with orbital, spin and isospin quantum numbers $k$, $s$ and $q$. 
The occupation factors $\eta_k^q$ are determined by the Fermi energy and 
the desired scheme of occupation (see discussion below). 
The kinetic energy and spin--orbit densities are defined by 
\begin{eqnarray}
\tau_q ( \boldsymbol{r} ) 		
	& = &	\sum_{k,s} \ \eta_k^q \, | \boldsymbol{\nabla} \varphi_k^q (\boldsymbol{r},s) | ^2 , \\
\boldsymbol{J}_q ( \boldsymbol{r} )	
	& = &   -i \sum_{k,s,s'} \ \eta_k^q \, (\varphi_k^q)^\ast (\boldsymbol{r},s') \
		\boldsymbol{\nabla} \varphi_k^{q} (\boldsymbol{r},s) 
		\times \langle s'| \boldsymbol{\sigma} | s \rangle\,.
\end{eqnarray}
The gradient of the spin--orbit density 
$\boldsymbol{\nabla} \boldsymbol{J} 
= \boldsymbol{\nabla} \boldsymbol{J}_p + \boldsymbol{\nabla} \boldsymbol{J}_n$ 
can be directly evaluated without 
first calculating $\boldsymbol{J}$:
\begin{equation}
\boldsymbol{\nabla} \boldsymbol{J}_q (\boldsymbol{r})
	=  -i \sum_{k,s,s'} \ \eta_k^q \, 
	   \boldsymbol{\nabla}(\varphi_k^q)^\ast (\boldsymbol{r},s') \,
	   \times \boldsymbol{\nabla} \varphi_k^{q} (\boldsymbol{r},s) 
	   \cdot \langle s'| \boldsymbol{\sigma} | s \rangle .
\end{equation}
We left out the spin--gradient term \cite{SP:Chabanat98},
which is cumbersome to evaluate in three-dimensional calculations numerically
and not very important.

The single-particle wave functions are determined as solutions of the
Hartree-Fock equations
\begin{equation}	
\left\{ - \boldsymbol{\nabla} \frac{\hbar^2 }{ 2m^{\ast}_q ( \boldsymbol{r}) } \boldsymbol{\nabla}
	+ U_q( \boldsymbol{r} )  
	-i\, \boldsymbol{W}_q( \boldsymbol{r} )\cdot (\boldsymbol{\nabla} \times \boldsymbol{\sigma}) 
	\right\}
	\varphi_k^q ( \boldsymbol{r} ) = \varepsilon_k^q\, \varphi_k^q (
	\boldsymbol{r}, s )\,,\label{eq:harfock}
\end{equation} 
with an effective mass term $m^{\ast} ( \boldsymbol{r})$, 
which depends on the $\mathcal{H}_{\text{eff}}$ part of the energy density functional
\begin{align}
\frac{\hbar^2}{2m_q^{\ast} ( \boldsymbol{r})} = \,
	& \frac{\hbar^2}{2m} 
	  + \textstyle{\frac{1}{8}} [ t_1 (2+x_1) + t_2 ( 2+ x_2)]\: 
	  \rho(\boldsymbol{r}) \notag \\
	& + \textstyle{\frac{1}{8}} [ t_2(1+2x_2) - t_1 (1 + 2x_1)]\: 
	  \rho_q (\boldsymbol{r} ), 
\end{align}
a nuclear central Potential
\begin{align}
U_q (\boldsymbol{r}) =\: 	
	& \textstyle{\frac{1}{2}} t_0 \big[ (2+x_0)\rho-(1+2x_0)\rho_q \big] \notag\\
	&+ \textstyle{\frac{1}{24}} t_3 (2+x_3)(2+\alpha)\rho^{\alpha +1} \notag \\
	&- \textstyle{\frac{1}{24}} t_3	(2 x_3+1) 
		   \big[ 2\rho^\alpha \rho_q 
			 + \alpha \rho^{\alpha -1}\big( \rho_p^2+\rho_n^2 \big) \big] \notag \\
	&+ \textstyle{\frac{1}{8}} \big[ t_1(2+x_1)+t_2 (2+x_2) \big] \, \tau \notag \\
	&+ \textstyle{\frac{1}{8}} \big[ t_2(2x_2+1)- t_1(2x_1+1) \big] \, \tau_q  \notag \\
	&+ \textstyle{\frac{1}{16}} \big[ t_2(2+x_2)-3t_1(2+x_1) \big] \, \Delta\rho  \notag \\
	&+ \textstyle{\frac{1}{16}} \big[ 3t_1(2x_1+1)+t2(2x_2+1) \big] \, \Delta\rho_q  \notag \\
	&- \textstyle{\frac{1}{2}} W_0 
	   \big[ \boldsymbol{\nabla}\boldsymbol{J} 
	   	 + \boldsymbol{\nabla} \boldsymbol{J}_q \big] \notag \\
	&+ \delta_{q,p} V_{\text{Coul}}
\end{align}
with the Coulomb field
\begin{equation}
V_{\text{Coul}}(\boldsymbol{r}) 
	= e^2 \int d^3r' \, \frac{\rho_C(\boldsymbol{r}') }{| \boldsymbol{r} - \boldsymbol{r'} | }
	\ - e^2 \left( \frac{3}{\pi} \right)^{1/3} \rho_C^{1/3} 
\end{equation}
and a spin-orbit field:
\begin{equation}
\boldsymbol{W}_q(\boldsymbol{r}) =\:	
	\textstyle{\frac{1}{2}} \, W_0 
	\left( \boldsymbol{\nabla} \rho + \boldsymbol{\nabla} \rho_q \right) 
\end{equation}

\subsection{Imaginary Time Step}
\label{imag_time_step}

Various different methods have been developed to solve the Hartree--Fock 
equations. 
Frequently the single-particle wave functions are expanded in a basis like e.g.
the eigenfunctions of an appropriate harmonic oscillator. This is appropriate
for describing the wave functions for single-particle states, which are deeply
bound. It is not so appropriate for the description of weakly bound or unbound
single-particle states, since the asymptotic behaviour of the harmonic 
oscillator basis states is not appropriate for these states. 

This can be cured by employing the eigenstates of a spherical box with an
appropriate radius $R$\cite{PC:Montani04}, which can also be considered as a
Wigner Seitz cell for describing periodic systems. Such a spherical box,
however, is not appropriate for the description of deformed nuclei and nuclear
structures as they are expected for the pasta phase in the crust of neutron
stars. This, as well as the problems with the boundary conditions in a spherical
WS cell discussed already in the introduction calls for a cartesian WS cell.
 
The single-particle wave functions in such a cartesian WS cell can be
represented by its values on a discretized mesh in this cell. The spacings
between the mesh points, $\Delta x$, $\Delta y$ and $\Delta z$ correspond to
truncations in momentum space. Smaller values for these spacings account for
larger momentum components in the wave functions. 
The obvious disadvantage of such calculations is the huge amount of mesh points
which has to be taken into account.
Therefore one needs a fast iterative procedure for the solution of the
self-consistent Hartree-Fock equations, which evaluates
only the desired states.

Davies et al. presented in \cite{IT:Davies80} an efficient method for this 
problem, 
the \emph{imaginary time step method}, 
which we want to outline briefly.
The origin for the name of this method is the analogy to the time--dependent 
Hartree-Fock (TDHF) method which solves the equations
\begin{equation}
 i \hbar \frac{\partial \varphi_k}{\partial t}  
 	= H(t) \varphi_k(t),  \ \ \ \ \ \ \ \ k = 1, \ldots , A
\end{equation}
for an orthonormal set of $A$ wave functions $\{ \varphi_k \}$, and a Hamiltonian $H$ 
which depends on the time $t$. This is the case when we identify $H(t)$ e.g.
with the Hartree-Fock (HF) Hamiltonian represented in eq.(\ref{eq:harfock}), which
depends on $t$ as it depends on the resulting wave-function $\varphi_k(t)$ in a
self-consistent way.  
These equations are discretized
in time introducing a time step $\Delta t$,
with $t_n = n\, \Delta t$. Then the time evolution of the set of wave functions
$\{ \varphi_k \}$ may be approximated by the iterative procedure
\begin{equation}
  |\varphi_k^{(n+1)} \rangle 
  	= \exp\left( - \frac{i}{\hbar} \Delta t \, H^{(n+\frac{1}{2})}  \right) 
	  | \varphi_k^{(n)} \rangle, \ \ \ \ \ \ \ k = 1, \ldots , A
\end{equation}
in which $\varphi_k^{(n)}$ represents the wave-function $\varphi_k$ at the time
$t_n$ and $H^{(n+\frac{1}{2})}$ denotes the numerical approximation to the Hamiltonian
$H(t)$ at the time $(n+ \frac{1}{2}) \Delta t$.  
The idea of Davies et al. was to replace the time step $\Delta t$ by 
the imaginary quantity $-i\Delta t $.
Introducing the positive parameter $\lambda = \Delta t / \hbar$
the procedure for the imaginary time step gets
\begin{equation}
  |\tilde{\varphi}_k^{(n+1)} \rangle 
  	= \exp\left( - \lambda \, H^{(n+\frac{1}{2})}  \right) 
	  | \varphi_k^{(n)} \rangle, \ \ \ \ \ \ \ k = 1, \ldots , A
\end{equation}
where $\{ \tilde{\varphi}_k^{(n+1)} \}$ is not any more an orthonormal set of wave functions
since the imaginary time operator 
$\exp \big( - \lambda \, H^{(n+\frac{1}{2})}  \big)$ is not unitary.
Applying the Gram--Schmidt orthonormalization method $\mathcal{O}$ we get the 
orthonormal set $\{ \varphi_k^{(n+1)} \}$ by
\begin{equation}
  | \varphi_k^{(n+1)} \rangle 
  	= \mathcal{O} | \tilde{\varphi}_k^{(n+1)} \rangle \ \ \ \ \ \ \ k = 1,
	\ldots , A\,.
\end{equation}
This procedure converges leading to those eigenfunctions of the Hamiltonian $H$,
which correspond to the lowest $A$ eigenvalues of the Hamiltonian $H$. 

In practical applications the Hamiltonian $H^{(n+\frac{1}{2})}$ is replaced
by the Hamiltonian $H^{(n)}$ of the $n$--th step, 
which makes the calculation
fast keeping the algorithm stable.
After this replacement, we get the following operation on the wave functions
\begin{equation}
  \varphi_k^{(n+1)} = 
	\mathcal{O} \left( \, \exp \big( - \lambda \, H^{(n)} \big) \ \varphi_k^{(n)} \, \right) 
	\ \ \ \ \ \ k=1, \ldots , A, 
\end{equation}
For numerical application one has to truncate the exponential series to a certain order. 
In earlier HF calculations the gradient method was used with the operation 
$ \mathcal{O} ( 1 - \lambda H ) $ on the wave functions (\cite{NMB:Ring80}). 
If one truncates the exponential series in the imaginary time step going beyond
the first order one obtains an improvement
of the gradient method.
Davies et al. recommended  a truncation to 4th or 5th order for a HF calculation
of $^{40}$Ca together with a
time step $\Delta t = 4.0 \times 10^{-24} \,\text{s}$ and a mesh size of $1.0 \, \text{fm}$.

In our calculations we used the same mesh size as Davies et al. but
the convergence got worse since we consider in our studies a larger number of
nucleons, which implies a larger number of wave
functions $A$ have to be evolved.  
Hence we truncated the exponential operator at 9th order and the time step $\Delta t$
was set to $2.0 \times 10^{-24} \, \text{s}$. 
For the check of convergence the mean square deviation of the single particle energies 
for $N$ Nucleons and $\eta_k$ the occupation probability is calculated by
\begin{equation}
 \Delta H^{(n)}
   = \left( \frac{1}{A} \sum_{k=1}^A \, \eta_k \, 
   	    \Big( \langle \varphi_k^{(n)} | H^{(n)^2} | \varphi_k^{(n)} \rangle
   	       	- \langle \varphi_k^{(n)} | H^{(n)} | \varphi_k^{(n)} \rangle^2  \Big)
	   \right)^{\frac{1}{2}} 
\end{equation}
which provides a better criterion as calculating energy differences.


The HF equations have been solved by discretization in coordinate space within
a cubic Wigner--Seitz cell similar to \cite{TD:Bonche85} with .
periodic boundary conditions. 
The box sizes typically considered vary from $2 \times 10$ fm to $2 \times 16$ fm.
This technique is able to allow for general deformations of the
quasi-nuclear structures. The densities we are considering requires to account
for around 1500 nucleons, which implies that up to $A=2600$ wave functions had
to be evolved to account for pairing correlations with occupation probabilities
$\eta_k$ different from zero. 

To decrease the numerical effort we assume two symmetries 
like in \cite{TD:Bonche85}:
\begin{itemize}
\item  parity
  \begin{equation}
  \hat{P}\varphi_k(\boldsymbol{r} , s ) 
  	= \varphi_k(-\boldsymbol{r}, s) 
	= p_k \varphi_k(\boldsymbol{r} , s ), 
	\ \ \ \ \ p_k = \pm 1;
  \end{equation}
\item z--signature 
  \begin{equation}
  \begin{split}
  \exp \{ i\pi ( \hat{J}_z - \textstyle{\frac{1}{2}} ) \}  \varphi_k(x, y, z, s ) 
  	&= \sigma \varphi_k ( -x, -y, z, s ) \\
	&= \eta_k \varphi_k ( x, y, z, s ),
	\ \ \ \ \ \eta_k = \pm 1. 
  \end{split}
  \end{equation}
\end{itemize}

These symmetries still allow triaxial deformations and
reduce the calculation to the positive coordinates in each direction.
As additional symmetry time--reversal--invariance is assumed for
the time--reversed pairs $\varphi_k$, and $\varphi_{\bar{k}}$:
\begin{equation}
  \varphi_{\bar{k}} ( \boldsymbol{r}, s ) 
  	= ( \hat{T} \varphi_k ) ( \boldsymbol{r}, s) 
	= \sigma \varphi^\ast_k ( \boldsymbol{r}, -s) .
\end{equation}
Summarising this symmetries it is sufficient to solve the 
HF equations for one wave function of the time--reversed pairs.
We choose the positive z--signature orbital for which
we get the symmetries summarised in  
table \ref{Skyrme_parity}. 
The wave functions $\varphi_k (\boldsymbol{r}, s)$ are realized as
complex Pauli spinors. 
The reflections at the $x=0$ and $y=0$ planes are realized 
by the parity operator of the real part together with complex conjugation.

\begin{table}
\begin{center}
\begin{tabular}{|c|ccc|}
\hline
\quad             & x=0   & y=0   & z=0  \\
\hline
Re $\varphi_k(\mathbf{r},+\textstyle{\frac{1}{2}})$  & $+$   & $+$   & $p_k$  \\
Im $\varphi_k(\mathbf{r},+\textstyle{\frac{1}{2}})$  & $-$   & $-$   & $p_k$  \\
Re $\varphi_k(\mathbf{r},-\textstyle{\frac{1}{2}})$  & $-$   & $+$   & $-p_k$  \\
Im $\varphi_k(\mathbf{r},-\textstyle{\frac{1}{2}})$  & $+$   & $-$   & $-p_k$  \\
\hline
\end{tabular}
\caption{ Parity properties of the Pauli spinors 
	  with respect to the coordinate planes }\label{Skyrme_parity}
\end{center}
\end{table}

The iteration is performed with accurate numerical methods. 
For the differential operators 11--point formulas are used, 
which have been derived by eliminating errors for functions $f$ with $f(x)=x^n$ 
up to a certain $n_0 \in \mathbb{N}.$ 
The ansatz for the numerical approximation of the derivatives 
on an equidistant mesh with the points $x_i$ and $f_i = f(x_i)$ is for the first derivative
\begin{equation}
\frac{\partial }{\partial x} f( x_i) 
	\approx \left( \frac{\partial }{\partial x} \right)_{\text{num}}  f( x_i)
	= \sum_{j=1}^N \, a_j \, \frac{1}{2j \Delta x} \, \left(  f_{i+j} - f_{i-j} \right),
\end{equation}
with $N=5$ for 11-point formula and $a_j$ the coefficients of the formula.
Requiring that the approximation gets equal up to a certain $n_0 \in \mathbb{N}$
we obtain a linear equation. Inserting the result in the ansatz we finally obtain
\begin{align}
&\left( \frac{\partial }{\partial x} \right)_{\text{num}}  f( x_i) \\
& =\frac{1}{\Delta x}  \Big( 
     \textstyle{\frac{1}{19860}} 
     ( 11 f_{i+5} -4500 f_{i+2} + 16350 f_{i+1}  \notag \\
    &\ \ \ \ \ \ \ \ \ \ \ \ \ \ \ \ \ \ \  - 16350 f_{i-1} + 4500 f_{i-2} -11 f_{i-5} ) \notag \\
    &\ \ \ \ \ \ \ \ \ \ \  +\textstyle{\frac{1}{55608}} 
       		( - 445 f_{i+4} + 2950 f_{i+3}  - 2950 f_{i-3} + 445 f_{i-4} )     
       		\Big).  \notag
\end{align}
For the second derivative used in the laplacian the ansatz is
\begin{equation}
\frac{\partial }{\partial x} f( x_i) 
	\approx \left( \frac{\partial }{\partial x} \right)_{\text{num}}  f( x_i)
	= \sum_{j=1}^N \, a_j \, \frac{1}{(j \, \Delta x)^2} \, \left(  f_{i+j} -2f_i + f_{i-j} \right).
\end{equation}
and finally the formula gets
\begin{align}
&\left( \frac{\partial^2 }{\partial x^2} \right)_{\text{num}}  f( x_i) \\
&=\frac{1}{(\Delta x)^2}  \Big( 
     \textstyle{\frac{1}{49650}} 
     ( 11 f_{i+5} -11250 f_{i+2} + 81750 f_{i+1}  \notag \\
&\ \ \ \ \ \ \ \ \ \ \ \ \ \ \ \ \ \ \ \ \ \ 
    		 + 81750 f_{i-1} - 11250 f_{i-2} + 11 f_{i-5} ) \,
		 - \textstyle{\frac{1729639}{595800}} f_i    \notag  \\
&\ \ \ \ \ \ \ \ \ \ \ \ \ \ \  
    		+\textstyle{\frac{1}{333648}} 
       		( - 1335 f_{i+4} + 11800 f_{i+3}  + 11800 2950 f_{i-3} - 1335 f_{i-4} )     
       		\Big).    \notag
\end{align}      

In the case of the Wigner--Seitz cell calculations
charge neutrality is assumed and
electrons are taken into account as relativistic Fermi gas which
contribute to the charge density  
$\rho_C(\boldsymbol{r}) = \rho_p(\boldsymbol{r}) - \rho_e $.
For the calculation of finite nuclei the electrons are not taken into account
($\rho_e = 0$).

There are different methods to solve the Poisson equation
\begin{equation}
 - \Delta V_C (\boldsymbol{r}) = 4\pi \rho_c (\boldsymbol{r}).
\end{equation}
It turned out that the numerically most accurate and stable 
method is the integration applying the Green's function for this problem
\begin{equation} 
  V_C(\boldsymbol{r}) 
	= \int_V dr'^3 \, \rho_C(\boldsymbol{r'})\, \frac{1}{|\boldsymbol{r}-\boldsymbol{r'}|}.  
\end{equation}
Unfortunately, this integral has lots of singularities, but it can be rewritten.
First, the Green's function is written as (\cite{Vautherin73}):
\begin{equation}
  \frac{1}{|\boldsymbol{r}-\boldsymbol{r'}|} 
	= {\textstyle \frac{1}{2}} \, \Delta_{\boldsymbol{r'}} |\boldsymbol{r}-\boldsymbol{r'}|.
\end{equation}
Then the integral is transformed by Green's theorem for scalar functions
$f$ and $g$ defined on a Volume $V$ with closed surface $A=\partial V$ (\cite{Jackson75}):
\begin{equation}
  \int_V dV \, (f\, \Delta g) - \int_V dV \, ( g \, \Delta f) 
	= \oint_{A=\partial V} \boldsymbol{dA} \cdot (f\, \boldsymbol{\nabla}g - g \, \boldsymbol{\nabla}f ).
\end{equation}
Identifying  and $f=\rho_C(\boldsymbol{r'})$ and 
$g = \frac{1}{2} |\boldsymbol{r}-\boldsymbol{r'}|$ the final result gets
\begin{equation}
\begin{split}
  V_C(\boldsymbol{r}) 
	=\ & \frac{1}{2} \int_V dr'^3 \, 
	 	\Delta\rho_C(\boldsymbol{r'})\, |\boldsymbol{r}-\boldsymbol{r'}|   \\
	  &+ \frac{1}{2} \oint_{A= \partial V} \boldsymbol{dA} \cdot 
	  \left( \rho_C(\boldsymbol{r'})\, 
		\boldsymbol{\nabla}_{\boldsymbol{r'}} |\boldsymbol{r}-\boldsymbol{r'}|
  		\, - \, |\boldsymbol{r}-\boldsymbol{r'}| \,
		\boldsymbol{\nabla}_{\boldsymbol{r'}} \rho_C(\boldsymbol{r'} ) 
	  \right),
\end{split}
\end{equation}
which has no singularities.
Altogether the result of this transformation behaves very well in numerical calculations 
and the numerical result is practically the same as the exact one. 
For finite nuclei it is possible to drop the boundary integrals as has
already been discussed by Vautherin \cite{Vautherin73}.

We tested the computer program for the parameter set Skyrme III by comparing
results for finite nuclei with those of  \cite{TD:Bonche85}. 
Additional tests have been performed using the parameter set SLy4 \cite{SP:Chabanat98}.

\section{Relativistic Mean Field Calculations}

In order to test the sensitivity of the results on the model under consideration
we also investigated the quasi-nuclear structures in the crust of neutron stars
employing the 
relativistic mean field approach in a cubic box. 

\subsection{From the Lagrangian to the Dirac Equation}
The relativistic mean field approach is based on a 
Lagrangian is similar to that in \cite{RMF:Reinhard89} and consists of
three parts: Lagrangian for the free baryons $\mathcal{L}_B$, 
the free mesons $\mathcal{L}_M$
and the interaction Lagrangian $\mathcal{L}_{\text{int}}$:
\begin{equation}
	\mathcal{L} = \mathcal{L}_B + \mathcal{L}_M + \mathcal{L}_{\text{int}},   
\end{equation}
which take the form
\begin{equation}
  \begin{split}
  \mathcal{L}_B =&  \bar{\Psi} ( \, i \gamma _\mu \partial^\mu - m ) \Psi,  \\
  \mathcal{L}_M =&	{\textstyle \frac{1}{2}}
			\Big( \partial_\mu \Phi_\sigma \partial^\mu 
			      \Phi_\sigma - m_\sigma^2 \Phi_\sigma^2 \Big)   \\
  		 &	- {\textstyle \frac{1}{2}} 
		        \sum_{\kappa = \omega, \rho, \gamma }
			\Big( {\textstyle \frac{1}{2}} F_{\mu \nu}^{(\kappa)}
				- m_\kappa^2 A_\mu^{(\kappa)} A^{(\kappa) \mu} \Big),        \\
  \mathcal{L}_{\text{int}} 
  		=&-\bar{\Psi} g_\sigma \Phi_\sigma \Psi 
		- \bar{\Psi} g_\omega \gamma_\mu A^{(\omega) \mu } \Psi     \\
		&- \bar{\Psi} {\textstyle\frac{1}{2}} g_\rho \gamma_\mu 
		  \boldsymbol{\tau } \boldsymbol{A}^{(\rho)\mu } \Psi 
                 - \bar{\Psi}e\gamma_\mu {\textstyle\frac{1}{2}}(1+ \tau_3 ) A^{(\gamma)\mu} \Psi ,              
  \end{split}
\end{equation}
with the field strength tensor
$F_{\mu \nu}^{(\kappa)} = \partial_\mu A_\nu^{(\kappa)} - \partial_\nu A_\mu^{(\kappa)} $,
the meson fields $ \Phi_\sigma$, $ A^{(\omega)} $, 
$ \boldsymbol{A}^{(\rho)} $ and the electromagnetic field $A^{(\gamma)}$.
The bold symbols are isovectors, the $\gamma^\mu$ are the Dirac $\gamma$ matrices
and $\Psi$ is a nucleon field which consists of Dirac 4-spinors with isospin space.
The masses are the baryon mass $m=938.9\, \text{MeV}$ and the meson masses   
$m_\sigma = 520 \, \text{Mev}$, $m_\omega= 783 \, \text{MeV}$ and $m_\rho= 770 \, \text{MeV}$
according to a parameter set for the linear model from Horowitz and Serot \cite{LHS:Horowitz81} 
cited as L-HS in \cite{RMF:Reinhard89}.
The coupling constants of this parameter set are $g_\sigma = 10.4814$, $g_\omega = 13.8144 $
and $g_\rho = 8.08488$. The charge of the electron $e = \sqrt{\alpha \, \hbar
c/4\pi}$ where $\alpha$
is the fine structure constant and $\hbar c = 197.32\, \text{MeV fm}$.
    
Applying the equations of motion and taking the static limit we obtain
in the Hartree approximation the static Dirac equation (\cite{Fritz94})
\begin{equation}\label{static Dirac}
\varepsilon_\alpha \, \psi_\alpha = (\boldsymbol{\alpha} \boldsymbol{p} + V + \beta (m - S) )\, \psi_\alpha.
\end{equation}
where $\alpha$ and $\beta$ are matrices like in \cite{Bjorken_Drell64}, 
$\varepsilon_\alpha$ the single-particle energy of the state $\psi_\alpha$, 
$\boldsymbol{p}$ the momentum operator and 
$S$ and $V$ the the scalar and vector fields
\begin{eqnarray}
S 	&=&	- g_\sigma \Phi_\sigma  \notag \\
V	&=&	g_\omega A_0^{(\omega)} + \textstyle{\frac{1}{2}} g_\rho \tau_3 A_0^{(\rho)}
		+ e\, \textstyle{\frac{1}{2}} (1-\tau_3)A_0^{(\gamma)}. 
\end{eqnarray}
For the mesons fields we get Klein-Gordan-equations.
After neglecting retardation effects and taking the
Hartree--approximation the meson field equations read
\begin{eqnarray}
  ( - \Delta + m_\sigma^2)\, \Phi_\sigma 	&=& 	-g_\sigma \, \rho^s  
  \notag\\
  ( - \Delta + m_\omega^2)\, A_0^{(\omega)} 	&=& 	g_\omega \, \rho \notag    \\
  ( - \Delta + m_\rho^2)\, A_0^{(\rho)} 	&=& 	{\textstyle \frac{1}{2}}
  g_\rho \, \rho_3 \notag\\
    - \Delta \, A_0^{(\gamma)} 		        &=& 	e \, \rho_C
    \label{eq:mesons}  
\end{eqnarray}
with the scalar density $\rho^s$, the baryon density $\rho$, the 
The densities are calculated taking into account only the 
occupied positive energy states in the Fermi--sea and neglecting the 
negative energy states in the Dirac--sea (''No-sea'' approximation)
\begin{eqnarray*}
  \rho^s	&=&	\sum_{\alpha=1}^N \eta_\alpha  \, \bar{\psi}_\alpha \psi_\alpha	\\
  \rho		&=&	\sum_{\alpha=1}^N \eta_\alpha  \, \bar{\psi}_\alpha \gamma_0 \psi_\alpha	\\
  \rho_3	&=&	\sum_{\alpha=1}^N \eta_\alpha  \, \bar{\psi}_\alpha \tau_3 \gamma_0 \psi_\alpha  \\
  \rho_C	&=&	\sum_{\alpha=1}^N \eta_\alpha  \, 
  			\bar{\psi}_\alpha \frac{1}{2}(1-\tau_3)\gamma_0 \, \psi_\alpha  \ \  (- \rho_e),
\end{eqnarray*}
where $\eta_\alpha$ are the occupation numbers determined by the BCS-formalism.
The electron density $\rho_e$ has been considered for the Wigner--Seitz cell calculations
but not for studies of finite nuclei.

\subsection{Solving the Triaxial Dirac Equation}

The solution of the Dirac equation for nucleons in a cubic box 
differs of course in some aspects from that in the spherical one.
Therefore we briefly outline the numerical solution in the following.

We decompose $\psi_\alpha$ in an upper and lower component Pauli spinor:
\begin{equation}
\psi_\alpha = \begin{pmatrix} \varphi_\alpha \\ \chi_\alpha  \end{pmatrix}
\end{equation}
Hence after applying the transformation 
\mbox{$\varepsilon_\alpha \rightarrow \varepsilon_\alpha - m$}
to energy levels without rest-mass 
the Dirac equation becomes 
\begin{eqnarray}
\varepsilon_\alpha \, \varphi_\alpha &=& 
	\boldsymbol{\sigma} \boldsymbol{p} \ \chi_\alpha
	+ U_\varphi \, \varphi_\alpha \notag	\\
\varepsilon_\alpha \, \chi_\alpha &=& 
	\boldsymbol{\sigma} \boldsymbol{p} \ \varphi_\alpha
	+ U_\chi \, \chi_\alpha
\end{eqnarray}
with the potentials
\begin{eqnarray}
	U_\varphi 	&=& 	- S + V    	\notag	\\
	U_\chi		&=&	- 2m + S + V.		
\end{eqnarray}
Now we obtain an ''effective Schroedinger equation'' 
by inserting the lower component into the equation for the upper one.
This method is according to Reinhard \cite{RMF:Reinhard89} the most efficient way to 
solve the Dirac equation.
First we modify the lower component
\begin{equation}
( \varepsilon_\alpha - U_\chi ) \, \chi_\alpha
	 = \boldsymbol{\sigma} \boldsymbol{p} \ \varphi_\alpha
\end{equation}
then we introduce an ''effective mass term'' which depends on the wave function
\begin{equation}
	\mathcal{B}_\alpha = \frac{1}{\varepsilon_\alpha - U_\chi }
\end{equation}
and finally get the ''effective Schroedinger equation''
\begin{equation}
	\varepsilon_\alpha \, \varphi_\alpha
	= \boldsymbol{\sigma} \boldsymbol{p} \, \mathcal{B}_\alpha \, 
		\boldsymbol{\sigma} \boldsymbol{p} \ \varphi_\alpha + U_\varphi \, \varphi_\alpha.
\end{equation}
So far the procedure corresponds to the method employed in calculations assuming
spherical symmetry\cite{MEW:Rufa88}. Using the discretization in a cartesian box,
however, requires a different treatment of angular momentum and spin--orbit
terms. With the help of the following formula for vector fields $\mathbf{A}$ and
$\mathbf{B}$  commuting with $\boldsymbol{\sigma}$
\begin{equation}
	\boldsymbol{\sigma} \mathbf{A}\, \boldsymbol{\sigma} \mathbf{B} 
 	=  \mathbf{A} \mathbf{B} + i \boldsymbol{\sigma} \, (\mathbf{A} \times \mathbf{B})
\end{equation}
we obtain from the relativistic kinetic energy term
\begin{equation}
	\boldsymbol{\sigma} \boldsymbol{p} \, \mathcal{B}_\alpha 
	\, \boldsymbol{\sigma} \boldsymbol{p} \, \varphi_\alpha
	= - \boldsymbol{\nabla} \mathcal{B}_\alpha 
	  \boldsymbol{\nabla} \varphi_\alpha
	  - i \, (\boldsymbol{\nabla} \mathcal{B}_\alpha)  
	    \cdot  ( \boldsymbol{\nabla} \times \boldsymbol{\sigma} ) \, \varphi_\alpha
\end{equation}
which is like the non-relativistic kinetic energy term plus the spin--orbit term for the upper component. 
From a further modification we obtain an expression called ''effective Hamiltonian'' ready for implementation
\begin{equation}
H_{\varphi,\alpha} \, \varphi_\alpha 	
	= - \mathcal{B}_\alpha \, \Delta \, \varphi_\alpha
	  - ( \boldsymbol{\nabla} \mathcal{B}_\alpha ) 
	     \cdot (\boldsymbol{\nabla} \varphi_\alpha) 
	  - i \, (\boldsymbol{\nabla} \mathcal{B}_\alpha) 
	     \cdot ( \boldsymbol{\nabla} \times \boldsymbol{\sigma}) \ \varphi_\alpha
	  + U_\varphi \ \varphi_\alpha.
\end{equation}
In order to calculate the eigenvalue $\varepsilon_\alpha$
 we can't use the ''effective Hamiltonian'' like a normal Hamiltonian 
because we have to take into account the effects of the lower component. 
This we do in the following manner:
\begin{equation}
  \chi_\alpha = \mathcal{B}_\alpha \, \boldsymbol{\sigma} \boldsymbol{p} \ \varphi_\alpha
\end{equation}
and hence the next approximation of the eigenvalue $\varepsilon_\alpha^{(n+1)}$ in
the iteration scheme gets
\begin{equation}
\varepsilon_\alpha^{(n+1)} = 	
	\int d^3r \,
	\left( \varphi_\alpha^{\ast} H_{\varphi,\alpha} \varphi_\alpha 
	      + \varepsilon_\alpha^{(n)} \chi_\alpha^{\ast}  \chi_\alpha \right)
\end{equation}
This means that in the lower component 
the whole new information is contained in the new Pauli spinor.
The total binding energy is calculated like in \cite{RNM:Walecka86}
using  cartesian coordinates 
\begin{equation}
\begin{split}
  E = 	& \sum_{\alpha} \eta_\alpha\, \varepsilon_\alpha 
    	  - \frac{1}{2} \int d^3r \, 
	  \Big( -g_\sigma \Phi_\sigma (\boldsymbol{r}) \rho^s(\boldsymbol{r})  \\
	&        +g_\omega A_0^{(\omega)} \rho (\boldsymbol{r}) 
		 + {\textstyle \frac{1}{2}} g_\rho A_0^{(\rho)} \rho_3(\boldsymbol{r})
		 + e A_0^{(\gamma)} \rho_C( \boldsymbol{r})
	  \Big)  \\
	& + E_{\text{cm}} + E_{\text{pair}}
\end{split}  
\end{equation}
with a
center of mass correction in the case of finite nuclei 
\begin{equation}
  E_{\text{cm}} = - {\textstyle \frac{3}{4}} \hbar \omega \ \ \
    \text{with} \ \ \hbar \omega = 41 \, A^{-1/3}\,\text{MeV}
\end{equation}
in compliance with \cite{DRH:Hofmann01} and a pairing energy 
$E_{\text{pair}}$ described in the next section.

For the variation of the wave functions in the cubic box
we employ once more the imaginary time step in the following manner:
First we operate on the upper component with the imaginary time step
and the ''effective Hamiltonian'':
\begin{equation}
	\varphi_\alpha^{(n+1)} = \exp (-\lambda H_{\varphi,\alpha} )\, \varphi_\alpha^{(n)}
\label{eq:relit1}\end{equation}
then the lower component is calculated:
\begin{equation}
  \chi_\alpha^{(n+1)} 
	= \mathcal{B}_\alpha\, 
	  \boldsymbol{\sigma}\boldsymbol{p}\, \varphi_\alpha^{(n+1)}
\label{eq:relit2}\end{equation}
and finally both components are orthonormalized together 
via the Gram--Schmidt method 
considering the symmetries of the Dirac spinors.
The symmetries are the same as in the Skyrme--Hartree--Fock calculations 
which are time reversal invariance, parity and z--signature.
These symmetries furthermore prevent the solution from ``slipping'' 
into the Dirac sea.
In case of a Dirac spinor these symmetries 
result in parity properties summarised in Table \ref{Dirac_parity} 
which corresponds 
to the Dirac spinor ansatz in spherical symmetry written in 
\cite{OP:Kleinmann94}.

\begin{table}[H]
\begin{center}
\begin{tabular}{|c|ccc|}
\hline
\quad             & x=0   & y=0   & z=0  \\
\hline
Re $\varphi_\alpha(\mathbf{r},+\frac{1}{2})$  & $+$   & $+$   & $p_\alpha$  \\
Im $\varphi_\alpha(\mathbf{r},+\frac{1}{2})$  & $-$   & $-$   & $p_\alpha$  \\
Re $\varphi_\alpha(\mathbf{r},-\frac{1}{2})$  & $-$   & $+$   & $-p_\alpha$  \\
Im $\varphi_\alpha(\mathbf{r},-\frac{1}{2})$  & $+$   & $-$   & $-p_\alpha$  \\
\hline
Re $\chi_\alpha(\mathbf{r},+\frac{1}{2})$  & $-$   & $-$   & $-p_\alpha$  \\
Im $\chi_\alpha(\mathbf{r},+\frac{1}{2})$  & $+$   & $+$   & $-p_\alpha$  \\
Re $\chi_\alpha(\mathbf{r},-\frac{1}{2})$  & $+$   & $-$   & $p_\alpha$  \\
Im $\chi_\alpha(\mathbf{r},-\frac{1}{2})$  & $-$   & $+$   & $p_\alpha$  \\
\hline
\end{tabular}
\caption{ Parity properties of the 
	  Dirac spinor with respect to the coordinate planes \label{Dirac_parity}}
\end{center}
\end{table}

For a comparison the energy of homogeneous
asymmetric nuclear matter is calculated similar to \cite{Avancini03}.

\subsection{Numerical Procedure}

The numerical method for solving the equations for the baryonic wave functions
(\ref{eq:relit1}) and (\ref{eq:relit2}) is essentially the same as in the case 
of the Skyrme Hartree--Fock approach.
Thus we restrict the discussion in this section to the solution of the meson 
field equations (\ref{eq:mesons}) and add some 
comments on the imaginary time step.

The meson equations have been solved with a finite difference scheme 
employing the conjugate gradient iterator operating on 
the meson fields with periodic boundary conditions.
The conjugate gradient iterator has the numerical advantage that there is
no operator matrix needed but only the operation of the differential operator 
on the meson field.  
The conjugate gradient method has been developed to solve linear 
equations \cite{CG:Hestenes52,NCG:Reid71}
and is now applied to a whole variety of numerical problems
for example to finite element solver 
for elliptic boundary value problems on an adaptive mesh 
with hierarchical basis preconditioning \cite{Yserentant86}, 
which provides a very fast algorithm.
The main idea of the conjugate gradient step is to solve the linear equation
$Ax-b =0$ with the linear operator $A$ and a vector $b$ by searching the 
minimum of the quadratic form 
\begin{equation}
  q(x) = {\textstyle \frac{1}{2}} x^T A x -b^Tx \,.
\end{equation}
In order to search the solution numerically one can use an iteration scheme
following the gradient method. 
Then it was discovered that the iteration
is accelerated if one searches not straight in gradient direction
but in the hyper-plane perpendicular to all previous directions. 
Theoretically the conjugate gradient step converges in less or equal
steps than the dimension of the vector space. In practical 
applications the machine errors require a restart after a certain 
amount of steps.

The overall numerical procedure has a good convergence.
In the imaginary time step the step $\Delta t$ for $\lambda = \Delta t/ \hbar$
could be set to $4.0 \times 10^{-24}\, \text{s}$ 
which is even larger compared to the corresponding Skyrme calculations.
In the test runs we obtained results with the parameter set L-HS
which agree with \cite{LHS:Horowitz81} 
within numerical accuracy.
We used this parameter set 
also for the actual calculations to compare the 
main properties of the 
relativistic mean field with the Skyrme calculations
in the Wigner--Seitz cell.

\section{Pairing Correlations}

Various properties of a neutron star, like e.g. its fluidity, the opacity with
respect to neutrino propagation etc., are very sensitive to occurrence of pairing
correlations. 
Therefore we included the possible effects of pairing in all calculations.
Our special attention was focussed on isospin $T=1$ pairing 
for nucleon pairs with total momentum equal to zero in the $^1S_0$ partial
wave like in an earlier approach in a spherical
box \cite{PC:Montani04}. 
Using the standard BCS approach the pairing gap 
$\Delta_k $ 
for pair of nucleons with momenta $k$ and $-k$ is obtained by solving the gap 
equation\cite{Kuckei03}
\begin{equation}
\Delta_k = - \frac{2}{\pi} \int_0^\infty dk' k'^2 V(k,k') 
	\frac{\Delta_{k'}}{2 \sqrt{(\varepsilon'_k - \varepsilon_F )^2 + \Delta_{k'}^2}} .
\end{equation}
Here $ V(k, k') $ denotes the matrix elements of the 
NN interaction in the $^1S_0$ partial wave, 
$\varepsilon_k$ the single particle energy for a nucleon 
with momentum $k$ and $\varepsilon_F$ the Fermi energy.

Instead of using the matrix elements of a realistic NN--interaction
which is fit to the scattering data we 
have decided to use the density dependent zero 
range effective interaction by Bertsch and Esbensen \cite{DP:Bertsch91}:
\begin{equation}\label{eff_pair}
V(\mathbf{r}_1, \mathbf{r}_2 ) 
	= V_0 \left( 1- \kappa \left( \frac{\rho(\mathbf{r}_1) }{\rho_0}\right)^\alpha \right) 
		\delta( \mathbf{r}_1 - \mathbf{r}_2 ) 
\end{equation}
with the parameters $V_0=481$ MeV fm$^3$, $\kappa = 0.7$, $\alpha = 0.45$
and the cut--off parameter for the gap-equation $ \varepsilon_c = 60$ MeV.
These parameters were derived from a realistic NN interactions by Garrido et al. 
\cite{PP:Garrido99}.

The occupation probabilities $\eta_k=v_k^2$ which are used to define the
densities of the Skyrme Hartree-Fock or the relativistic mean field approach
are determined from the 
quasi-particle energies $E_k$ \cite{NMB:Ring80}
\begin{eqnarray}\label{BCS_occ}
  v_k^2 &=& \frac{1}{2} \left( 1 - \frac{ \varepsilon_k - \varepsilon_F }{E_k} \right) \\
  u_k^2 &=& \frac{1}{2} \left( 1 + \frac{ \varepsilon_k - \varepsilon_F }{E_k} \right)
\end{eqnarray}
with
\begin{equation}     
  E_k = \sqrt{ (\varepsilon_k - \varepsilon_F)^2 + \Delta_k^2 },
\end{equation}
in which the pairing gap $\Delta_k$, the single particle energy $\varepsilon_k$
and the Fermi energy $\varepsilon_F$ enters.
The BCS--equations have to be solved in a 
self--consistent procedure fixing the Fermi energy $\varepsilon_F$
by the particle number condition for $N$ nucleons:
\begin{equation}
  N = \sum_k v_k^2\,.
\end{equation}

From the coefficients $u_k$ and $v_k$ of 
the standard BCS approach \cite{NMB:Ring80}
and the corresponding single--particle wave functions 
$\varphi_k$ one can calculate the anomalous density
\begin{equation}\label{eq:anomaldens}
  \chi (\mathbf{r}) = {\textstyle \frac{1}{2}} 
  			\sum_k \, u_k v_k \left| \varphi_k(\mathbf{r}) \right|^2.
\end{equation}
For a zero range pairing interaction as the one of eq.(\ref{eff_pair}), a
a local gap function can be defined:
\begin{equation}
  \Delta (\mathbf{r}) = - V(\mathbf{r}) \, \chi(\mathbf{r})\,.\label{eq:locald}
\end{equation}

The pairing correlations for continuous asymmetric nuclear matter have been
evaluated using the techniques described in \cite{Kuckei03,PNM:Fayans00}. 

\section{Results and Discussions}

In the first part of this section we are going to discuss the results of
Hartree-Fock calculations using the Skyrme force with the parameter set SLy4 as
defined in \cite{SP:Chabanat98}. The calculations are performed in a Wigner
Seitz (WS) cell with a shape of a cubic box. The size of the box $R$ has been assumed
to be identical in all 3 cartesian directions and has been adjusted to minimise
the total energy per nucleon for the density under consideration. The
calculations have been performed for charge neutral matter containing protons,
electrons and neutrons in $\beta$--equilibrium.

\begin{figure}
\begin{center}
\mbox{  \includegraphics[width = 11cm]{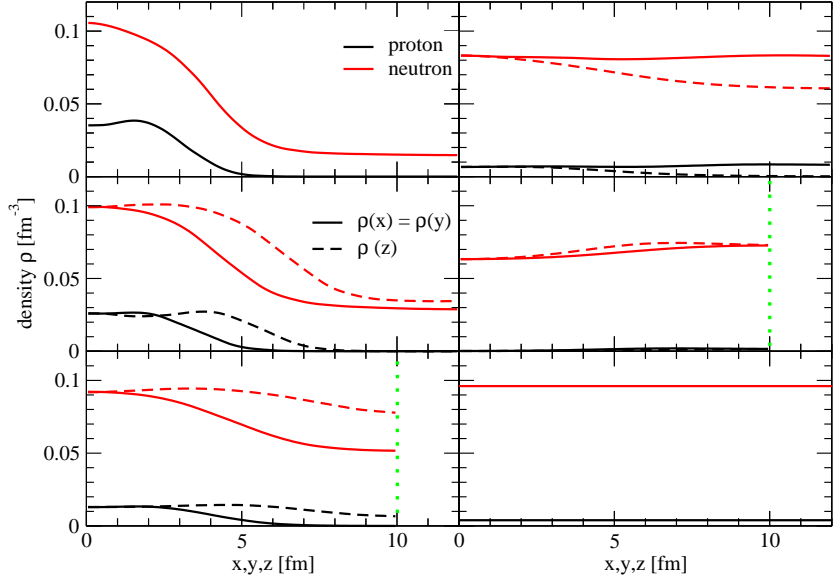}  }
\end{center}
\caption{\label{fig:shap1} (Color online) 
Density distributions resulting from Skyrme HF
calculations for protons (black color) and neutrons (red color) as a function 
of cartesian coordinates $x,y,z$. The panels in the left column refer to
densities 0.0166 fm$^{-3}$ (top), 0.0317 fm$^{-3}$, and 0.0565 fm$^{-3}$
(bottom), while those in the right column are obtained for baryon densities
0.0681 fm$^{-3}$ (top), 0.079 fm$^{-3}$, and 0.1 fm$^{-3}$. Further discussion
in the text.}
\end{figure}

A few typical density distributions resulting from these variational
calculations are displayed in Fig.~\ref{fig:shap1} with densities increasing
from top to bottom and all densities displayed in the left column being larger
than those in the right part of the figure.   

We start our discussion with the top panel in the left column representing a
nuclear structure at a baryonic density of 0.0166 fm$^{-3}$. In this case the
density profiles are identical in all 3 cartesian directions. This means that we
obtain a quasi-nuclear structure with spherical symmetry in the center of the WS
cell. The proton density drops to zero at a radial distance of around 4 fm.  The
neutron density profile drops around the same radius from a central density of
around 0.1 fm$^{-3}$ to the peripheral value of around 0.01 fm$^{-3}$. This
means that at this density we have obtained a structure of quasi-nuclear
droplets forming a cubic lattice, which is embedded in a sea of neutrons.

The second panel in the left part of Fig.~\ref{fig:shap1} displays the density
distributions, which have been obtained at a density of 0.0317 fm$^{-3}$. In
this case we obtain deformed quasi-nuclear droplets with radii, which are
slightly larger in one direction (chosen to be the $z$-direction, dashed curves)
than in the other two, which means that we find prolate deformation. 

\begin{figure}
\begin{center}
\mbox{  \includegraphics[width =7cm]{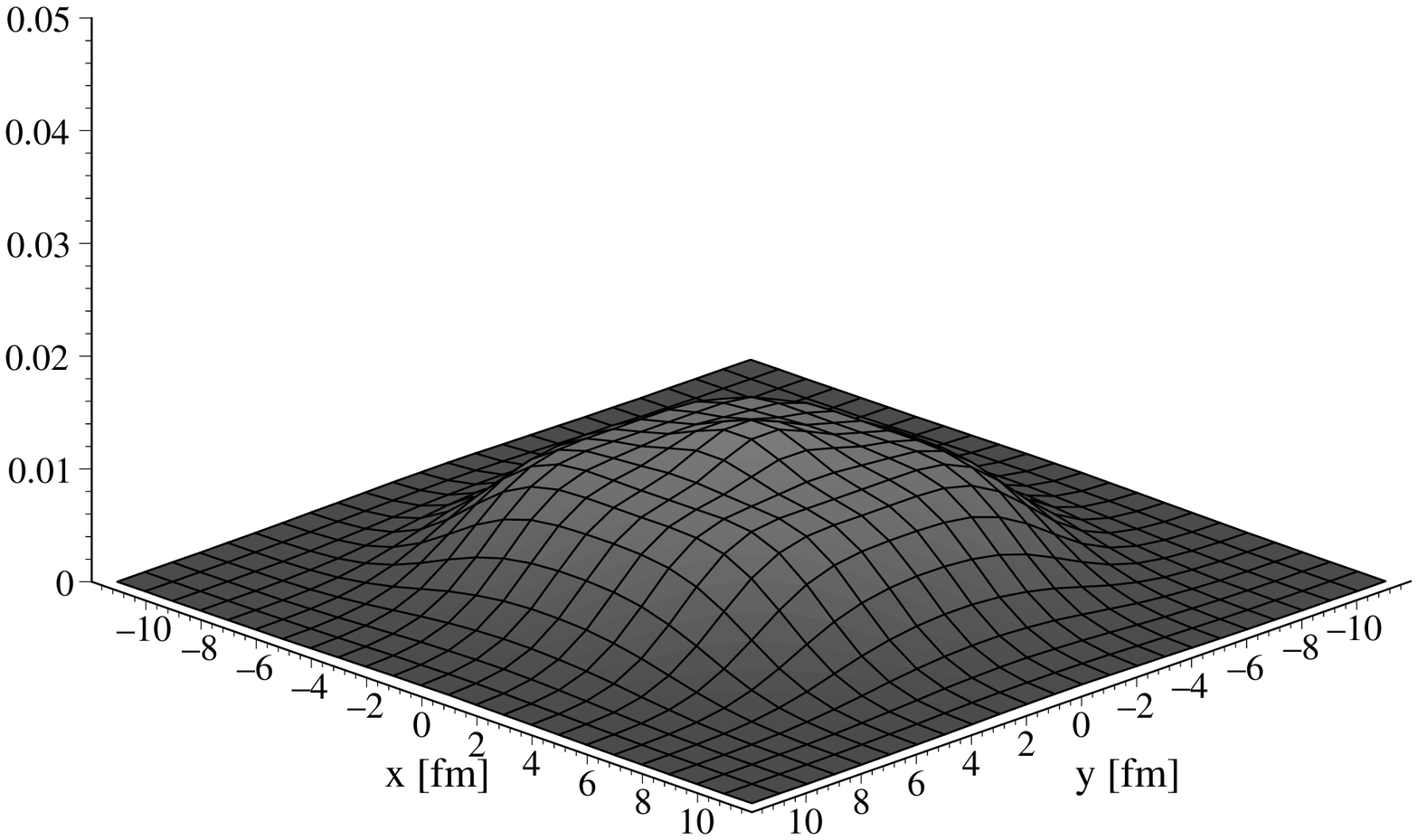} 
\includegraphics [width =7cm]{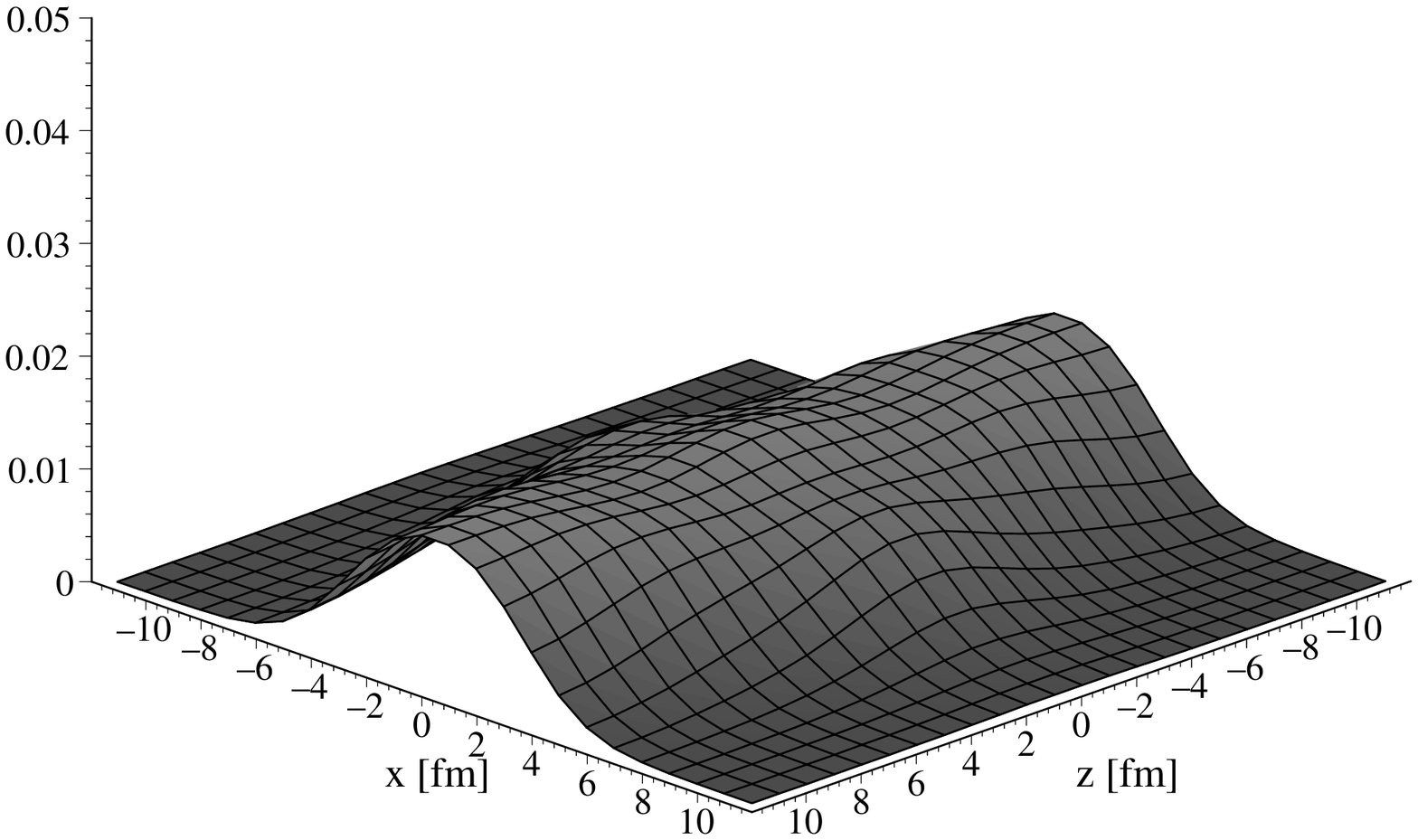}}
\end{center}
\caption{\label{fig:prof1} Profiles for the proton density distribution forming
a rod-structure at a density of 0.0625 fm$^{-3}$.}
\end{figure}

At slightly larger densities the deformation of the quasi-nuclear structures
increase until we reach a density at which the proton density does not vanish
along one of the three axis. Such an example (baryon density 0.0565 fm$^{-3}$)
is displayed in the bottom panel of the left column. In this case we have
quasi-nuclear structures in the shape of rods parallel to the z-axis. The
density of these rods is not homogeneous along the symmetry axes. Note that in
this example size of the WS cell became so small (R=10 fm) that the distance
from the center to the boundary of the WS box lies within the range displayed in
this figure and therefore has the boundary been indicated by the dotted line in
this panel. This structure is also displayed in Fig.~\ref{fig:prof1}, where the
profile of the proton-density is displayed in the xy and xz plane, respectively.

\begin{figure}
\begin{center}
\mbox{  \includegraphics[width =7cm]{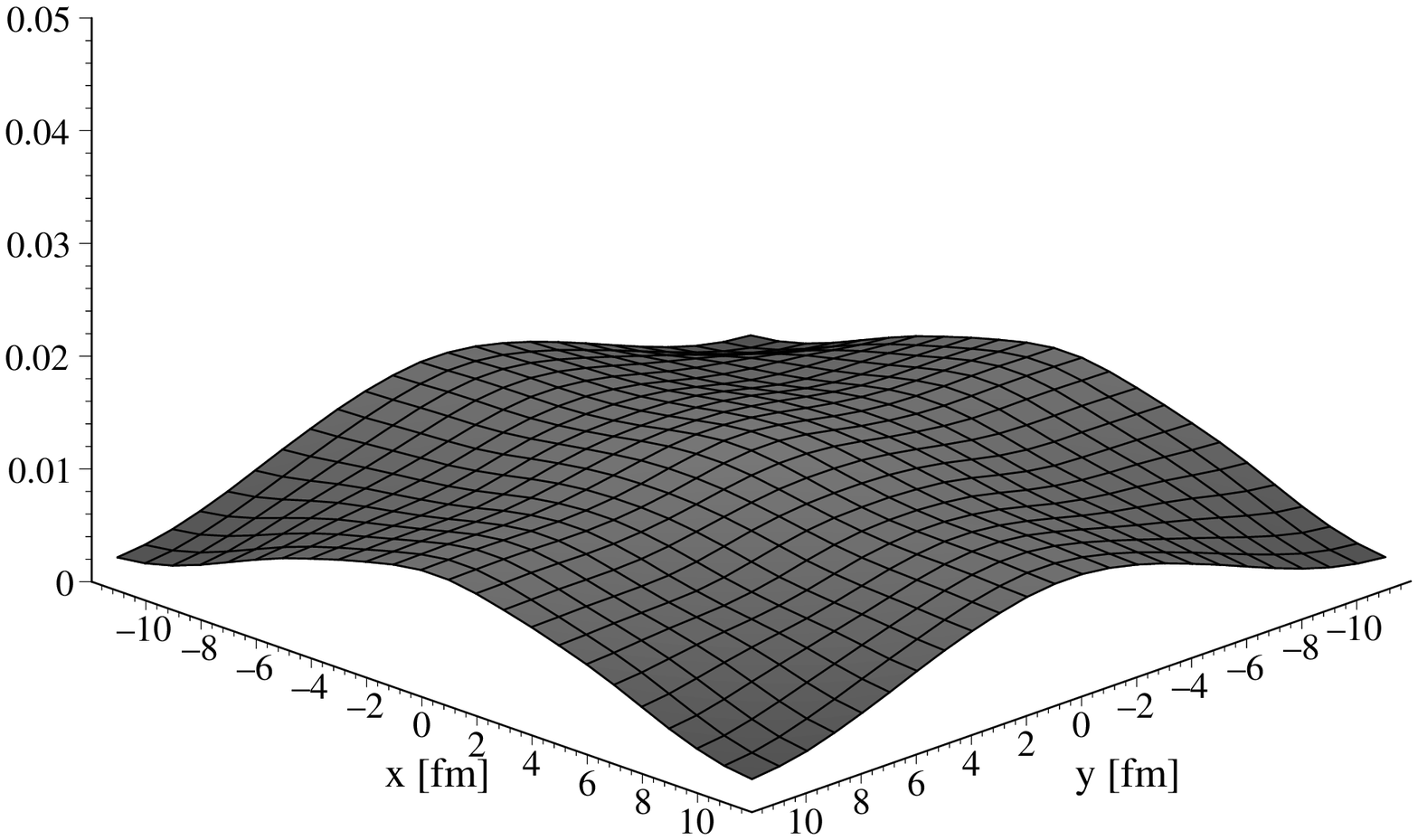} 
\includegraphics [width =7cm]{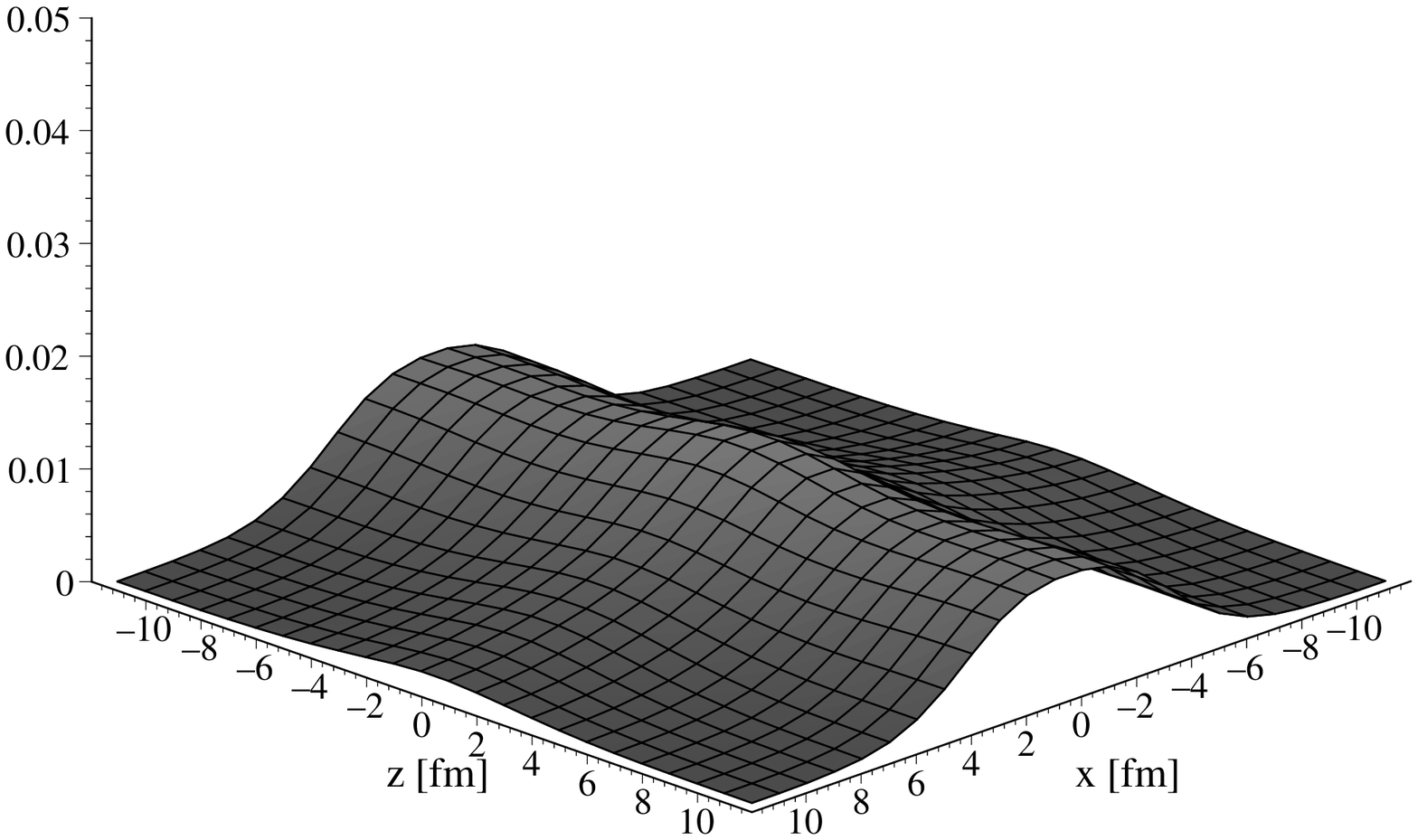}}
\end{center}
\caption{\label{fig:prof2} Profiles for the proton density distribution forming
a slab-structure at a density of 0.0775 fm$^{-3}$.}
\end{figure}

Performing HF calculations at a density of 0.0681 fm$^{-3}$ led to a density
density distribution as displayed in the top panel of the right column in
Fig.~\ref{fig:shap1}. In this example the proton as well as the neutron density
is essentially constant in the ($x,y,z=0$) plane. As a function of the third
coordinate ($z$, dashed lines) the proton density is reduced from the central
value at $z=0$ to zero at the border of the WS cell and also the neutron density
is reduced by about 25 percent going from the central to the peripheral values
of $z$. Therefore in this case we observe a structure in form of parallel slabs.
This slab structure is also displayed in Fig.~\ref{fig:prof2}. From this
presentation in particular it gets obvious that the density within such a slab
at $z=0$ is not really a constant but drops in particular along the diagonals
of the WS cell with $x=y, z=0$.
 
At even larger densities the Skyrme Hartree-Fock calculations in a cubic WS cell
yield structures, with smaller neutron densities in the center of the WS cell as
compared to the boundaries. An example of such an inverse structure,  which
corresponds to bubbles in the sea of nuclear matter, is displayed in the second
panel of the right column of Fig.~\ref{fig:shap1} at a density of 0.079
fm$^{-3}$. The proton density, which is hardly visible in this example, drops
from a peripheral value of around 0.004 fm$^{-3}$ to a central value of zero.

As a final example we present in the bottom panel of the right column of
Fig.~\ref{fig:shap1} the results of the HF calculation at a baryonic density of
0.1 fm$^{-3}$. At this and larger densities, the variational calculation yields
homogeneous nuclear matter in $\beta$-equilibrium. This example also demonstrates
that the cartesian box allows for a clean representation of the limit of
homogeneous matter. This is in contrast to calculations employing a spherical WS
cell. Depending on the boundary conditions used, calculations within such a
spherical box can lead to density profiles, which either show a maximum or a
minimum at the boundary. Even if one tries to use a set of boundary conditions,
which minimise this effect, the resulting density profile does not correspond to
the homogeneous solution \cite{PC:Montani04}. 

From this discussion we see that the HF calculations in a cartesian WS cell for
densities in the range of 0.01 fm$^{-3}$ to 0.1 fm$^{-3}$ leads to quite a
variety of shapes and quasi-nuclear structures with smooth transitions in
between. Following the discussions above these structures may be characterised
as quasi-nuclei, rod-structures, slab structures (all embedded in a sea of
neutrons) and, finally, the homogeneous matter. The densities at which the
transitions from one shape to other occur according to our Skyrme HF
calculations are listed in table~\ref{tab1}. The transition densities are very
similar to those obtained in \cite{CNS:Magierski02}.

\begin{table}
\begin{center}
\begin{tabular}{|c|c|c|c|c|}
\hline
& \multicolumn{2}{|c|}{Skyrme} 
  & \multicolumn{2}{|c|}{RMF} \\
\quad             	& HF   	 & TF      & H       & TF  \\
\hline
droplet--rod   		& 0.042  & 0.066   & 0.070   & 0.062 \\
rod--slab		& 0.070  & 0.078   & -    & -  \\
slab--homogeneous		& 0.080  & 0.085   & 0.075    & 0.072  \\
\hline
\end{tabular}
\end{center}
\caption{\label{tab1}Comparison of densities at which shape transitions occur using the 
Skyrme and Relativistic Mean Field (RMF) approach. Results are compared,
employing the microscopic Hartree-Fock (HF), Hartree (H) or the Thomas-Fermi (TF) approach.
All entries are presented in fm$^{-3}$.}
\end{table}

\begin{figure}
\begin{center}
\mbox{  \includegraphics[width = 9cm]{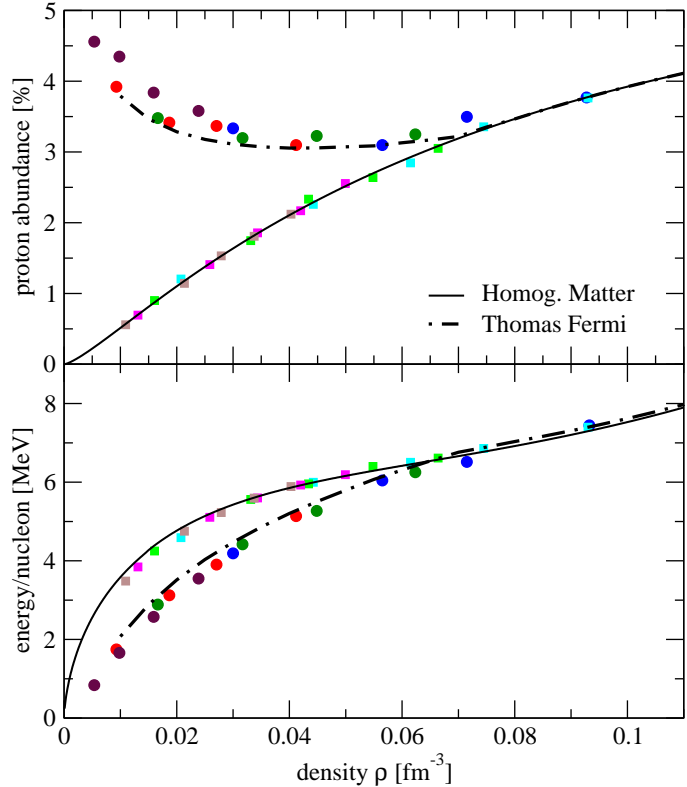}  }
\end{center}
\caption{\label{fig:sk-lead} (Color online) 
Proton-abundances and energy per nucleon as obtained from Skyrme Hartree-Fock
calculations at different densities. The results evaluated in cubic Wigner Seitz
cells (various symbols) are compared to those of homogeneous infinite matter
(solid lines) and of Thomas-Fermi calculations. Further details are given 
in the text.}
\end{figure}

The energies per nucleon and the proton abundances resulting from Skyrme
Hartree-Fock calculations are are displayed in the lower and upper panel of
Fig.~\ref{fig:sk-lead}, respectively. The solid lines indicate the results for
the evaluation of homogeneous matter in $\beta$-equilibrium. The results of 
calculations performed in cubic WS cells are presented in terms of individual 
symbols. Those symbols, which scatter around the homogeneous matter results are
obtained from WS calculations, constraining the HF single-particle wave
functions to plane waves. Therefore the scattering of these homogeneous matter
calculations within WS cells of finite size around the homogeneous result for
infinite matter is a measure of the shell-effects in the WS calculations on 
the calculated energy and proton abundances.

The Hartree-Fock calculations, which allow for the formation of inhomogeneous
quasi-nuclear structures, lead to a reduction of the calculated energy of 1 to 
2 MeV per nucleon. This gain in energy is reduced with increasing density up to
the density of 0.085 fm$^{-3}$ at which the energies of the inhomogeneous
structures merge into the results for the homogeneous matter. At densities
below this value of 0.085 fm$^{-3}$ the balance between the gain in binding
energy due to a local increase of the baryon density and the loss of binding
energy due to the localisation of nucleons and surface effects favours the
occurrence of inhomogeneities in the baryon densities. 

This balance between bulk energy arising from the energy density of nuclear
matter treated in a local-density approximation and surface effects is also
contained in the Thomas-Fermi (TF) approach. In this section we want to investigate
to which extent the results of our Hartree-Fock calculations can be reproduced
by corresponding  TF calculations. For that purpose we consider 
simple parametrisations for the density distribution for protons and neutrons,
which contain a constant peripheral density $\rho_q^{out}$ ($q=p$ or $n$ for
protons and neutrons, respectively) and an inner part describing the density
distribution in the center of the WS cell. For spherical quasi-nuclear
structures we employ the parametrisation of \cite{oyama2}
\begin{equation}
\rho_q(r) = 
    \begin{cases}
	(\rho_q^{in} - \rho_q^{out} )
	\left[ 1 - \left( \frac{r}{R_q} \right) ^{t_q} \right]^3 + \rho_q^{out}, & r < R_i \\
	\rho_q^{out}, & R_q \leq r\,.
    \end{cases}    \label{eq:param1}
\end{equation}
As an alternative we also consider a
Wood-Saxon density parametrisation of the form
\begin{equation}
\rho_q(r) = ( \rho_q^{in} - \rho_q^{out} ) 
	    \left[ 1+\exp \left( \frac{r - r_q}{a_q} \right) \right]^{-1} 
	    + \rho_q^{out}\,.\label{eq:paraws}
\end{equation}
For the description of rod-shape quasi-nuclear structures we use cylindrical
coordinates and parametrise the dependence of the densities on the radial
coordinate in a way corresponding to eqs.(\ref{eq:param1}) or (\ref{eq:paraws}).
In the case of quasi-nuclear structures in form of slab-shapes these
parametrisations are considered for the dependence of the densities on the
cartesian coordinate $z$.

Assuming those density distributions, the TF energy is calculated as a sum of
the bulk-energy, i.e. the integrated nuclear-matter energy densities, plus the
contribution of a surface term of the form \cite{oyama1,oyama2}
\begin{equation}
E_{\text{surf}} = F_0\,\int_{\text{WS-cell}} d^3r\,\left|
\boldsymbol{\nabla}\rho\right|^2 \,.\label{eq:esurf}
\end{equation}
The parameters of the density distributions in(\ref{eq:param1}) and 
(\ref{eq:paraws}) are varied to minimise the energy of the system under
consideration. The Parameter $F_0$ for the surface energy term in
(\ref{eq:esurf}) has been adjusted in two different ways. In a first approach we
have considered the properties of the nucleus $^{208}Pb$ and adjusted $F_0$ in
such a way that the TF calculation reproduced the energy and radius of this
nucleus derived from Skyrme HF. This leads to a value of $F_0$ of 68.3 MeV
fm$^5$  and 59.7 MeV fm$^5$ using the parametrisation of eq.(\ref{eq:param1})
and the Wood-Saxon parametrisation of eq.(\ref{eq:paraws}), respectively.

Adjusting the surface parameter $F_0$ in this way, one can evaluate the energies
of quasi-nuclear structures in a WS cell using the TF approximation. The results
for these TF energies are presented by the dashed dotted line in the lower panel
of Fig.~\ref{fig:sk-lead}. One finds that this procedure leads to energies,
which are consistently larger than those obtained in the HF calculations. It
seems that the TF approach, as it is used here, is underestimating the gain in
energy due to the formation of inhomogeneous structures. This could be a general
problem of the TF approximation or a result of the limitation in the variational
ansatz for the density functions. 

\begin{figure}
\begin{center}
\mbox{  \includegraphics[width = 9cm]{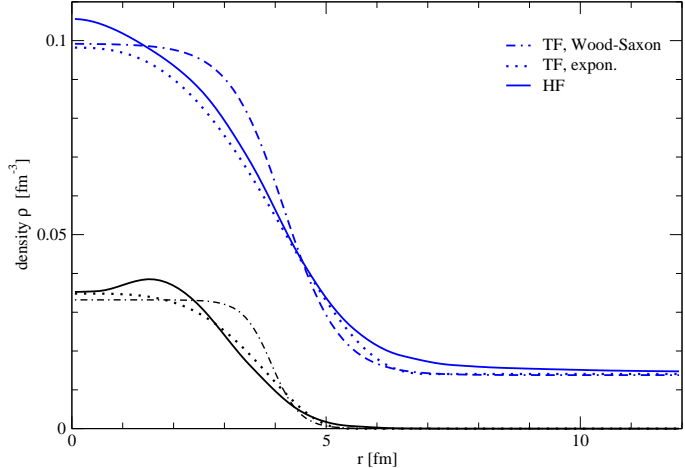}  }
\end{center}
\caption{\label{fig:tfdistri} (Color online) 
Density distributions resulting from Skyrme HF
calculations for protons (black color) and neutrons (blue color) as a function
of the distance from the center of the Wigner Seitz cell. The densities
resulting from HF are compared to those determined in Thomas-Fermi (TF)
calculations, assuming the parametrisation of (\protect{\ref{eq:param1}}),
dotted line, and (\protect{\ref{eq:paraws}}), dashed line. The example refers to
a global baryon density of 0.0166 fm$^{-3}$.
}
\end{figure}

To investigate these possibilities we have
considered the different parametrisations displayed in eqs.(\ref{eq:param1}) and
(\ref{eq:paraws}). It turns out that these two parametrisations lead indeed to
different density distributions, as displayed in the example of
Fig.~\ref{fig:tfdistri}, but it turns out that the resulting energy predictions
do not exhibit significant differences, so that we present only one example for
the TF approach in Fig.~\ref{fig:sk-lead}.

We then readjusted the the surface term in (\ref{eq:esurf}) to
 obtain an optimal fit of the HF energies for the quasi-nuclear structures in
$\beta$-equilibrium. This readjustment of the surface term leads to values of
the surface parameter $F_0$, which are about a factor of one half smaller than 
obtained from the fit to the properties of $^{208}Pb$. Using these readjusted
surface parameter we observe critical densities for the shape transitions of the
quasi-nuclear structures from droplets to rods to slabs and to homogeneous
nuclear matter at values which are similar to the results obtained in the HF
calculations. If, however, one uses
this reduced values derived from the fit to inhomogeneous matter in
$\beta$-equilibrium the TF calculation do not give an accurate description of
Hartree-Fock energies, in which the proton abundance has been fixed e.g. to a
value of 10 percent. This result can be taken as an indication that in addition
to the iso-scalar surface term of (\ref{eq:esurf}) an iso-vector surface term
might be required in addition to obtain a reliable TF approximation to the
results of corresponding HF calculations over a wide range of proton-neutron
asymmetries.

The upper panel of Fig.~\ref{fig:sk-lead} contains results on the proton
abundances for baryonic matter plus electrons in $\beta$-equilibrium. The value
of the proton abundance assuming homogeneous matter increases with density
reaching a value of about 4 percent at a baryonic density of 0.1 fm$^{-3}$.
Allowing for inhomogeneous, however, this value is almost constant around 3.2
percent in the density interval from 0.03 to 0.08 fm$^{-3}$ and yields even
larger values for densities below 0.03fm$^{-3}$. This trend is also reproduced
in the TF calculations. The increase of the proton abundances at smaller global
densities reflects the fact that at those small densities we observe local
structures in the center of the WS cells, with large local densities. The proton
abundance in these quasi-nuclear droplets is significantly larger than the proton
abundance in the homogeneous matter with the same global density.
The scattering of the results for the proton abundances
as a function of density resulting from the HF calculations reflects the
shell-effects, which preferentially yield quasi-nuclear with closed shells for
the protons.

\begin{figure}
\begin{center}
\mbox{  \includegraphics[width = 9cm]{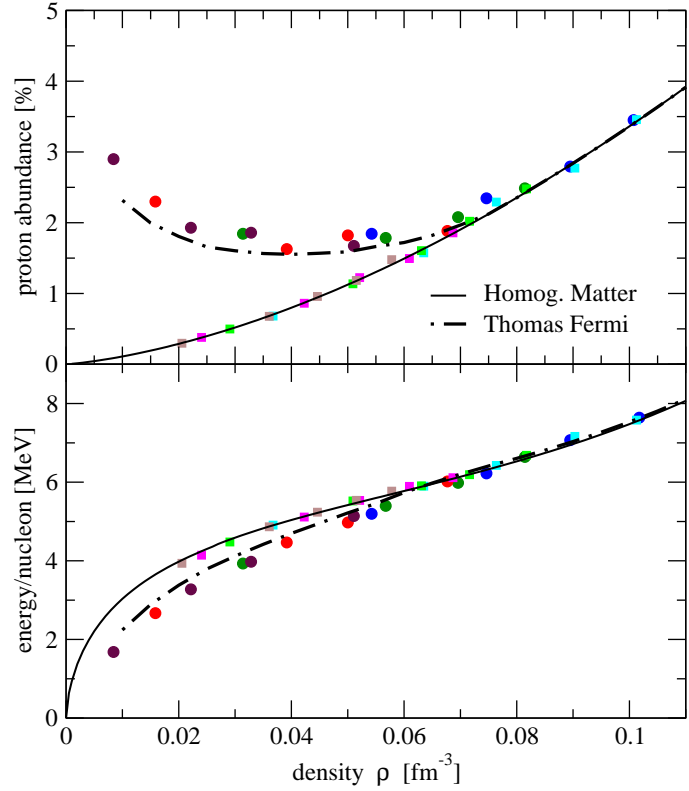}  }
\end{center}
\caption{\label{fig:rh-lead} (Color online) 
Proton-abundances and energy per nucleon as obtained from relativistic
mean-field
calculations at different densities. The results evaluated in cubic Wigner Seitz
cells (various symbols) are compared to those of homogeneous infinite matter
(solid lines) and of Thomas-Fermi calculations. Further details are given 
in the text.}
\end{figure}

A comparison of energies resulting from relativistic mean field calculations in
a Wigner Seitz cell are displayed in the lower panel of Fig.~\ref{fig:rh-lead}.
Comparing these results with the corresponding values displayed in 
Fig.~\ref{fig:sk-lead} one finds that the energy gain due to the formation of
inhomogeneous structures is much weaker in the relativistic mean field
calculations as compared to the Skyrme model. This is also reflected in the
corresponding Thomas-Fermi calculations. Note that also in this case we have
adjusted the constant $F_0$ of the surface term in (\ref{eq:esurf}) to reproduce
the bulk properties of $^{208}$Pb as predicted by the relativistic mean field
calculations. This leads to value for $F_0$  of 87.4 MeV
fm$^5$  and 80.3 MeV fm$^5$ using the parametrisation of eq.(\ref{eq:param1})
and the Wood-Saxon parametrisation of eq.(\ref{eq:paraws}), respectively. Both
values are significantly larger than the values required for $F_0$ in the case
of the Skyrme model used above.

The different interplay between volume-, surface-, symmetry- and Coulomb effects 
in the relativistic mean field model as compared to the Skyrme model also leads
to smaller values for the proton abundance in the region of nuclear densities, 
in which inhomogeneous structures emerge. The values around $\rho$ = 0.02
fm${-3}$, displayed in the upper panel of Fig.~\ref{fig:rh-lead}, are about 40
percent smaller than the corresponding values obtained in the Skyrme model (see
Fig.~\ref{fig:sk-lead}).
The differences in the balance between volume- and surface-contributions to the
energy also lead to different quasi-nuclear structures in the nuclear models
under consideration. It is worth mentioning that within the relativistic mean
field mode we do not find any formation of slab-like structures. Therefore the
table~\ref{tab1} contains for this case only transition densities for droplet to
rod structures and the formation of a homogeneous structure.

\begin{figure}
\begin{center}
\mbox{  \includegraphics[width = 9cm]{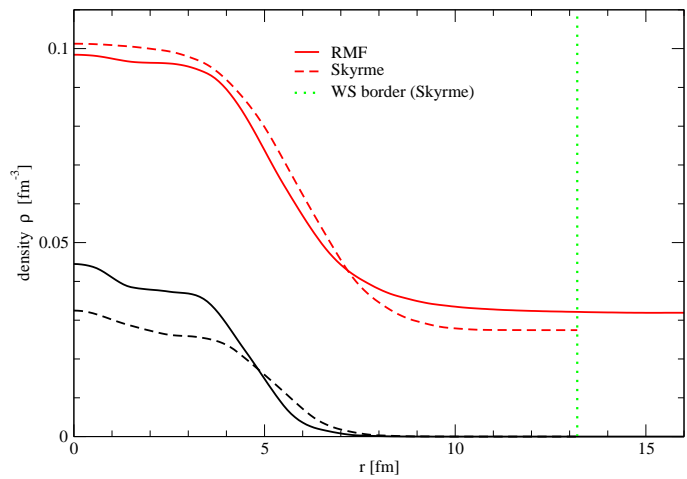}  }
\end{center}
\caption{\label{fig:dens_comp} (Color online) 
Density profiles for protons and neutrons as derived from Skyrme HF and
relativistic mean field calculations at a global density of $\rho$ = 0.032 
fm$^{-3}$.}
\end{figure}

The density profiles obtained from these 2 approaches also yield different
results. As an example we present in
Fig.~\ref{fig:dens_comp} the density profiles at $\rho$ = 0.032 fm$^{-3}$, a
density at which both the relativistic as well as the Skyrme model yield a
droplet structure. Note, that in the case of the Skyrme calculation we obtain a
Wigner Seitz cell with a length of 26.4 fm which leads to a borderline as
indicated by the dotted line, while the corresponding borderline for the RMF
calculation is identical to the frame of the figure. 

\begin{figure}
\begin{center}
\mbox{  \includegraphics[width = 11cm]{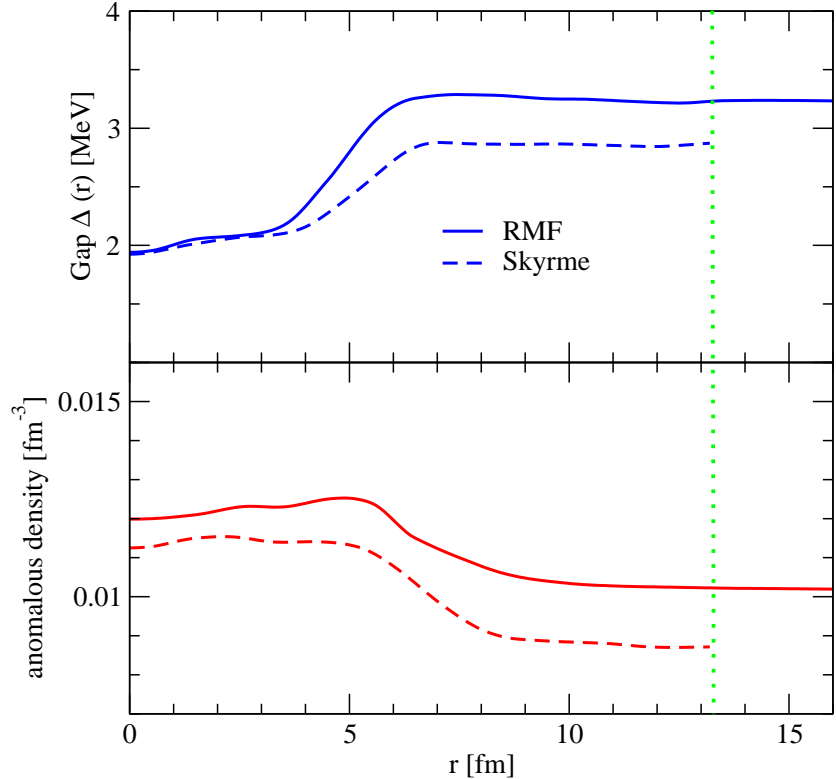}  }
\end{center}
\caption{\label{fig:pair} (Color online) Local pairing gap $\Delta (r)$ (upper
panel, see eq.(\protect{\ref{eq:locald}})) and anomalous density $\chi(r)$
(lower panel, see eq.(\protect{\ref{eq:anomaldens}})) for the configurations,
which are also considered in Fig.~\protect{\ref{fig:dens_comp}}.
}
\end{figure}

Finally, a feature of the pairing correlations obtained in these calculations
shall be discussed. For that purpose we present in the upper panel of
Fig.~\ref{fig:pair} the local pairing gap $\Delta (r)$ (see eq.(\ref{eq:locald})
for the formation of neutron pairs, as obtained in the Skyrme and relativistic
mean field model at $\rho$ = 0.032 fm$^{-3}$. In both of these approaches one
observes a suppression of the local gap $\Delta (r)$ in the region of the
quasi-nuclear structure, i.e. in the region where the density is large.

This phenomenon has been observed
before\cite{PC:Montani04,CNS:Magierski02,magierski07,baldo06} and has lead to  
discussions about various phenomena, which are related to to this periodic
structure of the gap parameter. It should be noted, however, that this
suppression of the gap parameter in the high-density region of the quasi-nuclear
structure is either to the local-density approximation, which is used to
calculate this local gap or to the assumption of the density-dependence of the
interaction strength for the pairing interaction, like the one, which we have
considered in our calculations (see eq.(\ref{eff_pair})). If, rather than
looking at the local gap parameter $\Delta (r)$, we inspect the anomalous
density $\chi(r)$ (see eq.(\ref{eq:anomaldens})), one finds even a small
enhancement of the anomalous density in the region of the quasi-nuclear structure.
This suggests that the reduction of the pairing gap in the region of high
densities might be an artefact of the special interaction considered. 

\section{Conclusions}

The structure of neutral baryonic matter is investigated in a region of
baryon densities between 0.01 and 0.1 fm$^{-3}$  performing
various Hartree-Fock an mean-field calculations with inclusion of pairing
correlations in a periodic lattice of Wigner-Seitz (WS) cells of cubic shapes. In 
this region of densities, which should occur in the crust of neutron stars, one
observes structures ranging from neutron-rich nuclei embedded in a sea of
neutrons up to homogeneous matter. The symmetries of the WS cell allow the
formation of triaxial structures but also include rod- and slab-like structures
and provide a natural transition to the description of homogeneous matter.    

For the baryonic components a Skyrme Hartree-Fock approximation has been
considered as well as a relativistic mean field model. Both approaches yield
an intriguing variety of quasi-nuclear structures with smooth transitions in
between. The occurrence of special structures as well as the critical densities
at which transitions between those structures occur depend on the nuclear model
considered. 

The resulting energies as well as the proton abundances can fairly
well be reproduced by a Thomas-Fermi approach, if the constant, determining the
strength of the surface term is adjusted to reproduce the results of the
microscopic calculations. A surface term depending on the isospin asymmetry
might be required to obtain Thomas-Fermi results, which are reliable over a
large interval of proton-neutron asymmetries.

Pairing-correlations have been evaluated within the BCS approach, assuming a 
density-dependent contact interaction. This leads to local pairing gaps for
neutron pairing, which
are significantly smaller in the regions of the quasi-nuclear structures as
compared to the bulk of the neutron sea. It is argued, however, that this
feature might be an artefact of the density-dependence of the effective pairing
interaction.

The present studies provide an interesting starting point for further studies on
the properties of matter in the crust of neutron stars. The single-particle
energies and wave-functions could be used for a microscopic study of
response-functions, which allow e.g. the evaluation of neutrino opacities.

This has been supported by the European Graduate School ``Hadrons in Vacuum in
Nuclei and Stars'' (Basel, Graz, T\"{u}bingen), which obtains financial support by
the DFG.




\end{document}